\documentstyle[12pt]{article}
\makeatletter
\newcommand{\singlespacing}{\let\CS=\@currsize\renewcommand{\baselinestretch}
{1.0}\tiny\CS}
\newcommand{\doublespacing}{\let\CS=\@currsize\renewcommand{\baselinestretch}
{1.5}\tiny\CS}

\newcommand{\ed}{\end{document}}
\newcommand{\bc}{\begin{center}}
\newcommand{\ec}{\end{center}}
\newcommand{\bfr}{\begin{flushright}}
\newcommand{\efr}{\end{flushright}}

\newcommand{\beq}{\begin{equation}}
\newcommand{\eeq}{\end{equation}}

\newcommand{\ben}{\begin{enumerate}}
\newcommand{\een}{\end{enumerate}}
\newcommand{\bit}{\begin{itemize}}
\newcommand{\eit}{\end{itemize}}

\newcommand{\ba}{\begin{array}}
\newcommand{\ea}{\end{array}}
\newcommand{\beqa}{\begin{eqnarray}}
\newcommand{\eeqa}{\end{eqnarray}}
\newcommand{\beqas}{\begin{eqnarray*}}
\newcommand{\eeqas}{\end{eqnarray*}}
\newcommand{\bfg}{\begin{figure}}
\newcommand{\efg}{\end{figure}}
\newcommand{\bds}{\begin{displaymath}}
\newcommand{\eds}{\end{displaymath}}
\newcommand{\btb}{\begin{tabbing}}
\newcommand{\etb}{\end{tabbing}}

\newcommand{\pad}{\partial}

\renewcommand{\a}{\alpha}
\renewcommand{\b}{\beta}
\newcommand{\g}{\gamma}

\newcommand{\eps}{\epsilon}
\newcommand{\th}{\theta}

\newcommand{\s}{\sigma}
\newcommand{\lam}{\lambda}

\newcommand{\pr}{\prime}

\newcommand{\m}{\mu}
\newcommand{\n}{\nu}

\newcommand{\p}{\pi}

\newcommand{\f}{\frac}

\begin{document}
\begin{center}{\large{\bf Magnus Force in High Temperature Superconductivity
and Berry Phase}}
\end{center}
\begin{center}
 D.Pal\
\footnote{e-mail : debjani$_{-}$p@hotmail.com} \\
 B. N. Mahavidyalaya, Itachuna, Hooghly\\
 B.Basu \footnote{ e-mail : banasri@www.isical.ac.in}\\
and\\
 P. Bandyopadhyay \footnote{e-mail :pratul@www.isical.ac.in}\\
 Physics and Applied Mathematics Unit\\
 Indian Statistical Institute\\
 Calcutta-700035
\end{center}
\date{}

\begin{abstract}
In the topological framework of high temperature
 superconductivity we have discussed the Magnus force acting on its vortices.\\
\end{abstract}
\begin{flushright}
\end{flushright}

In recent times, the debate on the problem of Magnus force
gained a renewed interest. There are two conflicting
points of view on the theory of transverse force.  Volovik [1]
has shown that the motion of the vortex with respect to the stationary
condensate induces a spectral flow. A momentum transfer from the
vortex system to a heat bath system is caused by a relaxation of the
quasiparticles of the vortex bound states (i.e., the electronic states inside
a vortex core). Therefore the vortex can apparently be moved without any
external source of transverse momentum. In this spectral flow theory the
coefficient of the transverse force $k$ essentially depends on the electronic
states inside a vortex core in combination of the relaxation time $\tau$ of
the quasiparticles. On the contrary, Ao and Thouless [2] showed that the
transverse force on a moving vortex is a robust quantity which does not depend
on the details of the vortex bound states inside a vortex core but only on the
superfluid density far from the core. Ao, and Thouless [2] calculated the
Berry phase for the adiabatic motion around a closed loop at zero temperature
and showed the existence of the Magnus force associated with Berry phase, as a
general property of vortex
line in a superconductor.

The purpose of the present note is to study the Magnus force in the vortex
dynamics of high $T_c$
superconductors. It is found to be consistent with the
idea of the Ao Thouless theory of the robust Magnus force. In some recent
papers [3,4] we have
 shown that
due to certain features in the background lattice how Berry's topological phase
plays an important role in describing high $T_c$ superconductivity.
Within this framework, we have studied here the Magnus
force required for a vortex to move.

In a recent paper [4], from a topological approach we have shown the relevance
of Berry phase in the
understanding of pairing mechanism in high $T_c$ superconductivity.
We know that the system of correlated electrons on a lattice is governed
by the Hubbard model which in the strong coupling limit and at half filling
can be mapped onto an antiferromagnetic Heisenberg model with nearest
neighbour interaction and is represented by the Hamiltonian
\beq
H = J \sum ( S^x_i S^x_j + S^y_i S^y_j + S^z_i S^z_j)
\eeq
with J>0.
For a frustrated spin system on a lattice
Wiegmann [5] has related two characterization operators of the ground state of
an
antiferromagnet, namely density of energy
\beq
\eps_{ij} = ( \f{1}{4} + \vec{S_i}. \vec{S_j})
\eeq
and chirality
\beq
W(C) = Tr \prod_{i \in C} ( \f{1}{2} + \vec{\s} . \vec{S_i})
\eeq
(where $\s$ are Pauli matrices and $C$ is a lattice contour) with the
amplitude and phase $\Delta_{ij}$ of
Anderson's resonating valence bond (RVB) through
\beq
\eps_{ij} = {| \Delta_{ij} |}^2
\eeq
and
\beq
W(C) = \prod_C \Delta_{ij}
\eeq
This suggests that $\Delta_{ij}$ is a gauge field. The topological order
parameter $W(C)$ acquires the form of a lattice Wilson loop
\beq
W(C) = e^{i \phi (c)}
\eeq
which is associated with the flux of the RVB field
\beq
e^{i \phi (c)} = \prod_C e^{i B_{ij}}
\eeq
$B_{ij}$ represents a magnetic flux which
penetrates through a surface enclosed by the contour $C$. This is
essentially the Berry phase related to chiral anomaly when we describe the
system in three dimensions through the relation
\beq
W(C) = e^{i 2 \pi  \m}
\eeq
where $\m$ appears to be a monopole strength.
In view of this, we consider a two dimensional frustrated spin system on a lattice residing on
the surface of a three dimensional sphere of a large radius in a radial
(monopole) magnetic field and associate chirality with the Berry
phase. In fact, the spherical geometry with a monopole at the center is
$equivalent$ to considering the effect of spin chirality in the RVB scenario
of the high temperature superconductors.

 In this geometry, we can
consider a generalised Heisenberg-Ising  Hamiltonian with nearest
neighbour interaction \beq H = J \sum ( S^x_i S^x_j + S^y_i S^y_j
+ \Delta S^z_i S^z_j) \eeq where $J > 0$ and the anisotropy
parameter $\Delta \ge 0$ and is given by $\Delta = \f{2\m + 1}{2}$
[6]. It is noted that $\m$ can take the values $\m = 0, \pm 1/2,
\pm 1, \pm 3/2 ........$. We observe that $\Delta = 1$ corresponds
to $\m = 1/2$. Indeed, the Ising part of the Hamiltonian
corresponds to the near neighbour repulsion caused by free
fermions and as $\m = 1/2$ is related to a free fermion, which
follows from the Dirac quantization condition $e\m=1/2$,the
condition $\Delta = 1$ gives rise to a isotropic
antiferrromagnetic Heisenberg model which is $SU(2)$ invariant.
When $\Delta = 0~ (\m = - 1/2)$ we have the $XX$ model.
 For a frustrated spin system, this effectively corresponds to a bosonic
system of spin singlets which  eventually
leads to a RVB state.

 To study the spinon and holon excitations, we consider a
single spin down electron at a site $j$ surrounded by an otherwise featureless
spin liquid representing a RVB state.
 We note that when the
single spin down state characterised by $\m=-\f{1}{2}$ is coupled with the
monopole in the background represented by $\m = -\f{1}{2}$ will give rise to a
state with $\m=-1$. Thus in this framework,
 the spinon is considered such that the
elementary spin 1 excitation characterized by $|\m|=1$ is split into two
parts, with one spin $\f{1}{2}$ excitation in the bulk and the other part
due to the {\it orbital spin} in the background characterized by the
chirality of a frustrated spin system.

It may be mentioned here that the spin singlet state forming the
quantum liquid are equivalent to FQH liquid with filling factor $\n =
1/2$ [3]. Indeed, in some recent papers [7-9] we considered a 2D electron gas of N-particles on the
surface of a three dimensional sphere in a radial (monopole)
strong magnetic field and  studied the behaviour of
quantum Hall fluid from the view-point of the Berry phase which is linked with
chiral anomaly . For the FQH liquid with even denominator filling factor
{\it i.e.} for the state with $\n = 1/2$, the
Dirac quantization condition $e \m = 1/2$ suggests that $\m=1$.
Then in the
angular momentum relation for the motion of a charged particle in the field of
a magnetic monopole
\beq
\vec{J} = \vec{r} \times \vec{p} - \m \vec{r}
\eeq
we note that for $\m = 1$ ( or an integer) we can use a transformation which
effectively suggests that we can have a relation of the form
\beq
\vec{J} = \vec{r} \times \vec{p} - \m \vec{r} = \vec{r^\pr} \times \vec{p^\pr}
\eeq
This indicates that the Berry phase which is associated with $\m$ may be
unitarily removed to the dynamical phase. This implies that the average
magnetic field may be taken to be vanishing in these states. However, the
effect of the Berry phase may be observed when the state is split into a pair
of electrons each with the constraint of representing the state $\m = \pm
1/2$. These pairs will give rise to the $SU(2)$ symmetry as we can consider
the state of these two electrons as a $SU(2)$ doublet. This doublet of Hall
particles
for $\n = 1/2$ FQH fluid may be taken to be equivalent to RVB singlets.

Now when a hole is introduced into the system by doping, the spinon will
interact with the hole through the propagation of the magnetic flux and this
coupling will lead to the creation of the holon which will have magnetic flux
$|\m_{eff}|=1$. Eventually, the residual spinon will be devoid of any magnetic
flux corresponding to $|\m_{eff}|=0$. This is realized when the single down
spin in the RVB liquid will form a pair with another up spin having $\m=+1/2$
associated with the hole following a spin pair. The holon having the effective
Berry phase factor $|\m_{eff}|=1$ will also eventually form a pair of holes
each having magnetic flux corresponding to $|\m|=1/2$ [4].

The Berry phase factor is
associated with the
chiral anomaly through the relation [10]
\beq
2 \m =- \frac{ 1}{2} \int \pad_\a J^5_\a d^4x
\eeq
where $J^5_\a$ is the axial vector current $\bar{\psi} \g_\a \g_5 \psi$.
When a chiral current
interacts with a gauge field we have the anomaly given by [11]
\beq
\pad_\a J^5_\a=- \frac{1}{8 \pi^2}Tr ~^\ast \tilde{F}_{\a\b} \tilde{F}_{\a\b}
\eeq
where $\tilde{F_{\a\b}}$ is the field strength associated with the gauge
potential $B_{\a}$ and from this we have the relation for the Pontryagin index
\beq
q~ =~2 \m = - \f{1}{16 \p^2} Tr \int~~^\ast \tilde{F}_{\a\b} \tilde{F}_{\a\b}
d^4 x
\eeq
In a frustrated spin system characterised by chirality this field can be
associated with the background magnetic field. Actually,
this gauge field is responsible for the spin-pairing and also for the hole
pairing. Due to this interacting magnetic fluxoid the hole pair can overcome
the bare Coulomb repulsion in high- $T_c$ superconductivity. The
superconducting phase order will be established when a spin pair with each
spin having
unit magnetic flux and a pair of holes with each hole having unit magnetic
flux interacts
with each other through a gauge force i.e., spin charge recombination comes
into play.
This helps us to infer of the topological aspect of pairing in
high $T_c$ superconductors and show that it is of magnetic origin [4].
We will show that this gauge field coupled with the vortex current will
lead to the transverse force responsible for the motion of the vortices.

It is known that a vortex line is topologically equivalent to a
magnetic flux. Thus in a cuprate superconductor the pair of charge carriers
each having magnetic flux associated with it may be viewd as a quantized
vortex line attached to each of them. These vortex lines lie along the
${\hat z}$ axis.
To study this vortex dynamics we assume $T=0$ and low magnetic field so that
vortex-vortex interaction can be ignored. To move a  vortex with
respect to the superconducting flow requires a transverse
lift force which
is known as the Magnus force. The Magnus
force acting on a vortex is proportional to the vector product of the velocity
of the vortex relative to the superconducting system and a vector directed
along the vortex core.

In our present formalism, we note that in the hole pair the associated flux
quantum
corresponding to $|\m|=1/2$ is derived from the bulk whereas the other flux
quantum with $|\m|=1/2$ is due to the background related to the chirality of
the frustrated spin system. In our model, we may assume that with the
movement of the hole pair, the associated vortex line corresponding to the
contribution from the bulk moves along with the centre of mass of the paired
charge carriers and the condensate will experience an interaction with the
background magnetic field. To study this interaction,
we have to introduce the $\th-term$ (last term in the Lagrangian (15)) as this
corresponds to the vortex line representing magnetic flux quantum associated
with the background magnetic field.
The Lagrangian
density of our model in spherical geometry, where the 2D surface is residing
on the surface of a 3D sphere of large radius with a monopole at the centre,
may
be written as
\beq
\begin{array}{lcl}
L~&=&\displaystyle{\frac{1}{2}[{\phi}^* ({\partial}_0-ieA_0) \phi-
\phi({\partial}_0+ieA_0) {\phi}^*] + \frac{1}{2m}{|({\partial}_a-ieA_a)
\phi|}^2 +\frac{\lambda}{2}{({|\phi|}^2-{\rho}_0)}^2+}\\
&&\displaystyle{~~~~~~~~~~~~~\frac{1}{4}F_{\a \b}F^{\a
\b}+\f{1}{4}^* \tilde{F}_{\a\b} \tilde{F}_{\a\b}}\\
\end{array}
\eeq
Here $\rho_0$ corresponds to the stationary configuration with
$|\phi|^2=\rho_0$. The term $F_{\a\b}$ corresponds to the electromagnetic
field strength and $\tilde{F}_{\a\b}$ corresponds to the background magnetic
field. $^* \tilde{F}_{\a\b}$ is the Hodge dual $$^* \tilde{F}_{\a\b}=\f{1}{2}
\epsilon_{\a\b\lam\s}F_{\lam\s}$$ It is noted that the P and T violating term
$^*\tilde{F}_{\a\b}\tilde{F}_{\a\b}$ takes care of the chirality of the system.
It is a
four divergence and hence does not contribute to the equation of motion but
quantum mechanically it contributes to the action. It is noted that there is a
singularity at the z-axis and hence we can take the two dimensional
formalism. To study the vortex dynamics, being inspired by Stone [12], we set
$\phi=fe^{i \th}$ so that we may write
\beq
L=if^2({\partial}_0 \th-ieA_0) + \frac{f^2}{2m}{({\partial}_a \th-ieA_a)
}^2 +\frac{\lambda}{2}{(f^2-{\rho}_0)}^2+\frac{1}{4}F_{\a \b}F^{\a
\b} + \f{k}{4 \pi} \epsilon_{\a\b\lam}B_{\a}{\partial}_{\b}B_{\lam}
\eeq
It is observed that the dimensional reduction suggests that the anomalous term
$^*
\tilde{F}_{\a\b}\tilde{F}_{\a\b}$ in 3+1 dimensions corresponds to the
Chern Simons
term $\epsilon_{\a\b\lam}B_{\a}{\partial}_{\b}B_{\lam}$ in 2+1 dimensions.
 We now
introduce Hubbard-Stratonovich fields $\vec{J}$ with the relation
$J_0=f^2$ to obtain \beq L \rightarrow
L^{\prime}=iJ_{\a}({\partial}_{\a} \th-ieA_{\a}) +
\f{1}{8mJ_0}{({\partial}_a
J_0)}^2+\f{m}{2J_0}J_a^2+\frac{\lambda}{2}{(f^2-{\rho}_0)}^2+gauge~
field~ terms \eeq We set the vortex part of the phase
$\th=\bar{\th}+\eta$ where $\bar{\th}=\arg(\vec{r}-\vec{r}_i(t))$
is the singular part of the phase due to vortices at $\vec{r_i}$
and $\eta$ is the non-singular part. Integration over $\eta$
suggests the conservation equation $\partial_{\a}J_{\a}=0$
indicating $J_{\a}$ as a current. So we can identify \beq J_{\a}=
\epsilon_{\a\b\lam}{\partial}_{\b}B_{\lam}=\f{1}{2}
\epsilon_{\a\b\lam} \tilde{F}_{\b\lam} \eeq such that the first
term in expression (17) corresponds to the interaction with the
background magnetic field. Indeed defining the vortex current \beq
K_{\a}=
\epsilon_{\a\b\lam}{\partial}_{\b}{\partial}_{\lam}\bar{\th} \eeq
we note that the first term in expression (17) can be written as
$$iB_{\a}(K_{\a}-e \epsilon_{\a\b\lam}{\partial}_{\b}A_{\lam})$$
This shows that
the
vortex current is coupled to the background gauge potential $B_{\a}$. It is
noted that $J_0$ has an equilibrium value $\rho_0$ even when the vortex is at
rest. Motion with respect to this background field gives rise to a Lorentz
force which is here just the Magnus force. So the
Magnus force is generated by the background magnetic field when it interacts
with the vortex current. In other words, the Magnus force is generated by the
background magnetic field associated with the chirality of the system.

To calculate this Magnus force we may take resort to the Berry phase approach
[2] .
When the vortex moves round a closed loop, we can express the Berry phase $e^{i
\phi}$ with
\beq
\phi~=~2 \pi N \m
\eeq
where $N$ is the total number of flux quantum enclosed by the loop. In our
approach each flux quantum in the background
is associated with a hole pair and so the number of flux quanta $N$
trapped is identical with the number of hole pairs enclosed by the loop.
Thus we can identify $N$ as the number of hole pairs and we can express $\phi$ as
\beq
\phi = 2 \pi \m \f{n_s}{2}
\eeq
where $n_s$ is the charged superfluid number density far from the vortex core.
The Magnus force is given by
the vector product of the vorticity and the motion relative to the
superconducting velocity
\beq
F_m~=~ \pm 2 \pi \f{n_s}{2} \m  \hat{c} \times \vec{V}_{vortex}
\eeq
Here +(-) corresponds to vortex parallel (antiparallel) to $\hat{c}$ axis and
 $\vec{V}_{vortex}$ is the velocity of
the vortex with respect to the superconducting velocity.
It is to be noted that the Magnus force explicitly depends on the number
of carriers instead of their mass. This supports the Ao, Thouless theory of
the origin of the Magnus force. As high $T_c$
superconductors are type-II superconductors, in the presence of an external
magnetic field, when some
magnetic flux quanta penetrates the material, the number density $n_s$ should
be replaced by $n$, the total density of the fluid when the radius of the
integration contour is much larger than the London penetration depth. This is
a consequence of the Meissner effect [13].

As we know, Aharanov- Casher phase is generated when the flux moves through
the mobile fluid charges. In the present situation, the phase arising out
of the flux moving through the fluid charges will be cancelled by that coming
from the flux motion through the static background ion charges. As the net
charge in the macroscopic region is zero, the two Aharnov- Casher phases will
cancel each other.

Here we can make a remark on the mysterios sign reversal of the
 Hall resistivity (conductivty) effect in the underdoped region in
cuprate superconductors [14].  It is
noted that in the underdoped region there will not be
sufficient number of holes to form superconducting pairs. So in this case
 a holon characterised by $|\m|=1$ will not be able to share the magnetic
flux with another hole and form the requisite pair. The integral value of $\m$
will lead to the removal of the Berry phase to the dynamical phase as given by
eqn.(11). Hence the Magnus force will be decreased. Besides, this in
combination with the magnetic flux lines induced by the
external magnetic field within the penetration depth may change the
orientation of the vortices. Indeed, the interaction of this single holon with
$\m=1$
with a magnetic flux line having $\m=- 1/2$(due to the external magnetic field)
 will correspond to $\m=1/2$ and as a result we will get a magnetic flux line
with
opposite orientation. This change in orientation of the magnetic flux
line will change the sign of the Hall conductivity. The change of the
electronic state due to doping could be related to the internal electronic
structure inside vortex core so that it affects the dynamic property of
vortices. Actually, some authors [15] have considered this many body effect
between
vortices and got results to support the Ao-Thouless theory. In our
field theoretical analysis through Berry phase we got the same result
by calculating the interaction of the background magnetic field with the vortex
current .


\vspace{2cm}

{\bf References}
\singlespacing
{
\begin{enumerate}
\item G. E. Volovik, J. Phys. C {\bf 20}, 1488 (1986).
\item P. Ao, D. L. Thouless,  Phys. Rev. Lett. {\bf 70}, (1993), 2158. \\
P. Ao, D. L. Thouless, Q. Niu, Phys. Rev. Lett. {\bf 76}, (1996), 758
\item  B. Basu, D.Pal and P. Bandyopadhyay, Int. J. Mod. Phys. {\bf B,
13} (1999), 3393.
\item  D.Pal and B. Basu, Europhys. Lett. {\bf 56}(1), 99, (2001).
\item P. Wiegmann, Prog. Theor. Phys. Suppl. {\bf 101}, 243, (1992)
\item P. Bandyopadhyay,  Int. J. Mod. Phys. {\bf A  15}, (2000), 1415.
\item  B. Basu and P. Bandyopadhyay, Int. J. Mod Phys. {B 11},
(1997), 2707.
\item  B.Basu, D.Banerjee and P.Bandyopadhyay, Phys.Lett. {\bf A 236}, (1997),
125.
\item B. Basu and P. Bandyopadhyay, Int. J. Mod Phys. {\bf B 12}. 2649,
(1998)
\item D.Banerjee and P.Bandyopadhyay, J.Math.Phys. {\bf 33}, (1992), 990.
\item A. Roy and P. Bandyopadhyay, J. Math. Phys. {\bf 30}, (1989), 2366.
\item M. Stone, Magnus And Other Forces On Vortices In Superfluids And
Superconductors : cond-mat/9708017.
\item M. R. Geller, C. Wexler and D. J. Thouless, Phys. Rev. {\bf B 57},
(1998), R8119.
\item T. Nagaoka et al., Phys. Rev. Lett. {\bf 80}(1998), 3504
\item N. Hayashi et.al.,   J. Phys. Soc. Jpn. {\bf 67},(1998), 3368
\end{enumerate}
}

\ed